\documentclass[aps,superscriptaddress,showpacs]{revtex4-1}

\usepackage{graphics}
\usepackage{epsfig}
\usepackage{dcolumn}
\usepackage{subfigure}
\usepackage{mathrsfs}
\usepackage{amsfonts}
\usepackage{amsmath}
\usepackage{amssymb}

\begin{document}

\title{Birman-Wenzl-Murakami Algebra, Topological parameter and Berry phase}

\author{Chengcheng Zhou}
\email{Zhoucc237@nenu.edu.cn}
\address{School of Physics, Northeast Normal University,
Changchun 130024, People's Republic of China}
\author{Kang Xue}
\email{Xuekang@nenu.edu.cn}
\address{School of Physics, Northeast Normal University,
Changchun 130024, People's Republic of China}
\author{Lidan Gou}
\address{School of Science, Changchun University of Science and Technology,
Changchun, 130022, People's Republic of China}
\address{School of Physics, Northeast Normal University,
Changchun 130024, People's Republic of China}
\author{Chunfang Sun}
\address{School of Physics, Northeast Normal University,
Changchun 130024, People's Republic of China}
\author{Gangcheng Wang}
\address{School of Physics, Northeast Normal University,
Changchun 130024, People's Republic of China}
\author{Taotao Hu}
\address{School of Physics, Northeast Normal University,
Changchun 130024, People's Republic of China}

\begin{abstract}
In this paper, a $3\times3$-matrix representation of Birman-Wenzl-Murakami(BWM) algebra has been presented. Based on which,
unitary matrices $A(\theta,\varphi_1,\varphi_2)$ and $B(\theta,\varphi_1,\varphi_2)$ are generated via
Yang-Baxterization approach. A Hamiltonian is constructed from the unitary $B(\theta,\varphi)$ matrix. Then we study Berry phase of
the Yang-Baxter system, and obtain the relationship between topological parameter and Berry phase.
\end{abstract}

 \maketitle

\vspace{0.3cm}

\section{Introduction}
To the best of our knowledge, the Yang-Baxter equation (YBE) was initiated in solving the one-dimensional $\delta$-interecting
models \cite{ye} and the statistical models \cite{be}. Braid algebra and Temperley-Lieb algebra (TLA) \cite{tla} have been widely used in the construction of YBE solutions \cite{kve,kk, yg,bob,bk,yql} and have been introduced to the field of quantum information, quantum computation and topological computation \cite{ayk,kl,frw,cxg1,sw,knot}. The Birman-Wenzl-Murakami (BWM) algebra \cite{bwma} which contain two subalgebra (Braid algebra and TLA) was first defined and independently studied by Birman, Wenzl and Murakami. Very recently \cite{tlaknot}, S.Abramsky demonstrate the connections from knot theory to logic and computation via quantum mechanics. But, the physical meaning of the topological parameter $d$ (describing the single loop in topology) is still unclear.

The geometrical phase \cite{b}, such as Berry phase, is an important concept in quantum mechanics \cite{aa,spe,sb,ts,w,bpxxz}.
In recent years, numerous works have been attributed to Berry phase \cite{bmk}, because of its possible applications to quantum computation (the so-called
geometric quantum computation) \cite{jve,dcz,ww,eeh}. Quantum logic gates based on geometric phases have been certified in both nuclear magnetic resonance \cite{jve} and ion trap based on quantum information architectures \cite{gpit}. The Ref. \cite{jve} pointed out geometric phases have potential fault tolerance when applied to quantum information processing. In 2007, Leek, P.J. \emph{et al.} \cite{bpss} illustrated the controlled accumulation of a geometric phase, Berry phase, in a superconducting qubit.

The Ref. \cite{hxg} applied TLA as a bridge to recast 4-dimensional YBE into its 2-dimensional counterpart. The 2-dimensional YBE have an important application value in topological quantum computation \cite{tqc1,tqc2}. To date, few studies have reported 3-dimensional YBE which may have potential application values in topological quantum computation. The motivation of this paper is twofold: one is that we structure 3-dimensional YBE , the other is to study the physical meaning of topological parameter $d$ from Berry phase. This paper is organized as follows: In Sec. 2, we introduce a specialized type BWM algebra, and present a \(3\times 3\)-matrix representation of BWM algebra. In Sec. 3, we obtain unitary matrices \(A(\theta,\varphi_1,\varphi_2), B(\theta,\varphi_1,\varphi_2)\) via Yang-Baxterization approach. Based on the solution, a Hamiltonian of the Yang-Baxter system is constructed, finally we study the Berry phase of this system. We end with a summary.

\section{BWM Algebra}
As we know the Braid relations are
\begin{equation}
\label{br}\left\{ \begin{aligned}
             &b_ib_{i\pm 1}b_i=b_{i\pm 1}b_ib_{i\pm 1},\\
             &b_ib_j=b_jb_i,\ |i-j|\geq 2,\\
\end{aligned} \right.
\end{equation}
where \(b_i=\stackrel{1}{I}\otimes \cdots \otimes \stackrel{i-1}{I}\otimes b\otimes \stackrel{i+2}{I}\otimes\cdots\).
When we just consider three tensor product space, the Braid relations becomes
\begin{equation}
b_{12}b_{23}b_{12}=b_{23}b_{12}b_{23},
\end{equation}
where \(b_{12}=b\otimes I, b_{23}=I\otimes b\), \(b\)-matrix is a \(N^2\times N^2\) matrix acted on
the tensor product space \(\nu \otimes \nu \), where \(N\) is the dimension of \(\nu\).
It is also well known that the braid relation can be reduced to a \(N\)-dimensional braid relation \((b_{12}\rightarrow A, b_{23}\rightarrow B)\)
\begin{equation}
\label{br1}ABA=BAB.
\end{equation}

Like this reduced method, we easily obtain \(N\)-dimensional reduced BWM-algebra relations from classical BWM-algebra relations.
The BWM algebra \cite{bwma,cgx,gx2,j} is generated by the unit $I$, the braid operators $S_i$ and the TLA operators $E_i$ and depends on two independent parameters $\omega$ and $\sigma$. Let us take the BWM relations as follows.
\begin{equation}
\label{bway} \left\{ \begin{aligned}
             &S_i-S_i^{-1}=\omega (I-E_i) ,\\
             &S_iS_{i\pm 1}S_i=S_{i\pm 1}S_iS_{i\pm 1},\ S_iS_j=S_jS_i,|i-j|\geq 2 ,\\
             &E_iE_{i\pm 1}E_i=E_i ,\ E_iE_j=E_jE_i ,\ |i-j|\geq 2 ,\\
             &E_iS_i=S_iE_i=\sigma E_i ,\\
             &S_{i\pm 1}S_iE_{i\pm 1}=E_iS_{i\pm 1}S_i=E_iE_{i\pm 1}
             ,\\
             &S_{i\pm 1}E_iS_{i\pm 1}=S_i^{-1}E_{i\pm 1}S_i^{-1} ,\\
             &E_{i\pm 1}E_iS_{i\pm 1}=E_{i\pm 1}S_i^{-1} ,\ S_{i\pm
             1}E_iE_{i\pm 1}=S_i^{-1}E_{i\pm 1} ,\\
             &E_iS_{i\pm 1}E_i=\sigma ^{-1}E_i ,\\
             &E_i^2=\left(1-\frac{\sigma-\sigma
             ^{-1}}{\omega}\right)E_i,
             \end{aligned} \right.
\end{equation}
where \(0\ne d=\left(1-\frac{\sigma-\sigma^{-1}}{\omega}\right)\in \mathbb{C}\) is a topological parameter in knot theory which does not depend on the sites of the lattices.

By reducing to the \(N\)-dimensional space \((S_{12}\rightarrow A ,
S_{23}\rightarrow B , E_{12}\rightarrow E_A , E_{23}\rightarrow
E_B)\), we have:
\begin{equation}
\label{bwa} \left\{ \begin{aligned}
             &A-A^{-1}=\omega (I-E_A) ,\ B-B^{-1}=\omega (I-E_B),\\
             &ABA=BAB ,\\
             &E_AE_BE_A=E_A ,\ E_BE_AE_B=E_B ,\\
             &E_AA=AE_A=\sigma E_A ,\ E_BB=BE_B=\sigma E_B ,\\
             &ABE_A=E_BAB=E_BE_A ,\ BAE_B=E_ABA=E_AE_B ,\\
             &AE_BA=B^{-1}E_AB^{-1} ,\ BE_AB=A^{-1}E_BA^{-1} ,\\
             &E_AE_BA=E_AB^{-1} ,\ E_BE_AB=E_BA^{-1} ,\\
             &AE_BE_A=B^{-1}E_A ,\ BE_AE_B=A^{-1}E_B ,\\
             &E_ABE_A=\sigma^{-1}E_A ,\ E_BAE_B=\sigma^{-1}E_B, \\
             &E_A^2=(1-\frac{\sigma-\sigma^{-1}}{\omega})E_A ,\
             E_B^2=(1-\frac{\sigma-\sigma^{-1}}{\omega})E_B,
      \end{aligned} \right.
\end{equation}
where \(A,B\) satisfy the \(N\)-dimensional Braid relation \eqref{br1}, \(E_A,E_B\) satisfy the \(N\)-dimensional TLA
relations
\begin{equation}
\label{tla} \left\{ \begin{aligned}
             &E_AE_BE_A=E_A ,\ E_BE_AE_B=E_B ,\\
             &E_A^2=d E_A ,\ E_B^2=d E_B ,
      \end{aligned} \right.
\end{equation}
It is interesting that Eq.\eqref{bway} and Eq.\eqref{bwa} have the same topological parameter \(d\).

In this paper, the \(A\)-matrix, \(B\)-matrix, \(E_a\)-matrix,
\(E_b\)-matrix, \(A(x)\)-matrix and \(B(x)\)-matrix are \(3\times
3\) matrices acting on the 3-dimensional space. To the TLA relations \eqref{tla}, we assume \(E_A\) and \(E_B\)
possess the same eigenvalues $d$ and 0. We assume $E_A$ is a diagonal matrix as following
\begin{equation}
E_A=\left(
\begin{array}{>{\displaystyle}l>{\displaystyle}c>{\displaystyle}r}
 0 & 0 & 0 \\
 0 & d & 0 \\
 0 & 0 & 0
\end{array}
\right).
\end{equation}
After tedious calculation,we obtain
\begin{equation}
E_B=\left(
\begin{array}{>{\displaystyle}l>{\displaystyle}c>{\displaystyle}r}
 \frac{d^2-d-1}{d } & \frac{\sqrt{d^2-d-1} }{d }e^{i \varphi_1} & -\frac{\sqrt{d^2-d-1} }{\sqrt{d} }e^{i (\varphi_1+\varphi_2)} \\
 \frac{\sqrt{d^2-d-1} }{d}e^{-i \varphi_1 } & \frac{1}{d} & -\frac{e^{i \varphi_2}}{\sqrt{d}} \\
 -\frac{\sqrt{d^2-d-1} }{\sqrt{d}}e^{-i (\varphi_1+\varphi_2)} & -\frac{e^{-i \varphi_2}}{\sqrt{d}} & 1
\end{array}
\right).
\end{equation}
It is worth to mention that \(E_B=UE_AU^{-1}\), and \(U\) is a unitary transformation matrix as follows
\begin{equation}
U=\left(
\begin{array}{>{\displaystyle}l>{\displaystyle}c>{\displaystyle}r}
 \frac{1}{(d-1)d } & -\frac{\sqrt{d^2-d-1}}{d} e^{i \varphi_1} & -\frac{ \sqrt{d^2-d-1} }{\sqrt{d} (d-1)} e^{i
(\varphi_1+\varphi_2)} \\
 \frac{\sqrt{d^2-d-1} }{d }e^{-i \varphi_1} & -\frac{1}{d} & \frac{e^{i \varphi_2}}{\sqrt{d}} \\
 \frac{ \sqrt{d^2-d-1} }{\sqrt{d} (d-1)}e^{-i (\varphi_1+\varphi_2)} & \frac{e^{-i \varphi_2}}{\sqrt{d}}
& -\frac{d-2}{d-1}
\end{array}
\right),
\end{equation}
where $d$, $\varphi_1$ and $\varphi_2$ are reals. The parameter $d$ is the so-called topological parameter.
For simplicity, we just consider the case of \(d>0\) in this paper.

The Ref. \cite{cgx} has explored \(S_i\) have 3 different
eigenvalues\((q,-q^{-1},q^{-2})\) in the BWM-algebra
(\emph{i.e.} Eq.\eqref{bway}). The same as \(E_A\) and \(E_B\), we assume \(A\) and \(B\) have the
same eigenvalues\((q,-q^{-1},q^{-2})\). The simplest \(A\) is
\begin{equation}
A=\left(
\begin{array}{>{\displaystyle}l>{\displaystyle}c>{\displaystyle}r}
 q & 0 & 0 \\
 \\
 0 & q^{-2} & 0 \\
 \\
 0 & 0 & -q^{-1}
\end{array}
\right),
\end{equation}
using the unitary transformation matrix \(U\), we have
\begin{equation}
B=UAU^{-1}=\left(
\begin{array}{>{\displaystyle}l>{\displaystyle}c>{\displaystyle}r}
 \frac{1}{q^4(d-1)d} & \frac{\sqrt{d^2-d-1} }{d q}e^{i \varphi_1} & -\frac{\sqrt{d^2-d-1} }{q^2(d-1) \sqrt{d}}e^{i (\varphi_1+\varphi_2)} \\
 \frac{\sqrt{d^2-d-1} }{d q}e^{-i \varphi_1} & \frac{q^2}{d} & \frac{q}{\sqrt{d}}e^{i \varphi_2} \\
 -\frac{\sqrt{d^2-d-1} }{q^2(d-1) \sqrt{d}}e^{-i (\varphi_1+\varphi_2)} & \frac{q}{\sqrt{d}}e^{-i \varphi_2} & \frac{d-2}{d-1}
\end{array}
\right),
\end{equation}
where \(d=q^{-1}+1+q\) and the parameter \( q \) is real. The matrices $A$ and $B$ satisfy the braid relation (\emph{i.e.} Eq\eqref{br1}). Towards braid relation, in some models \(\varphi_i, (i=1,2) \), may have a physical significance of magnetic flux. In the paper\cite{cxg1}, it has been shown the parameters \(\varphi_i\)'s are related to Berry phase.

Then we can verify that \(\{I,A,E_A,B,E_B\}\) satisfy the reduced BWM-algebra (\emph{i.e.} Eq.\eqref{bwa}), with \(d=q^{-1}+1+q\). Here we have set \(\omega=q-q^{-1}\)and \(\sigma =q^{-2}\). It is interesting that \(A,B,E_A,E_B\) are Hermitian matrices, and have the same similar transformation \(B=UAU^{-1}, E_B=UE_AU^{-1}\), where \(U\) is unitary (\emph{i.e.} \(U^{\dag}=U^{-1}\)).

\section{Yang-Baxterization, Hamiltonian, Berry phase }
In this section, A Hamiltonian is constructed from the unitary $B(\theta,\varphi)$ matrix. Then we study the Berry phase of the Yang-Baxter system, and obtain the relationship between the topological parameter and the Berry phase. We first explain the basic formula of YBE. The Yang-Baxter matrix $ \check{R} $ is a
$N^2\times N^2$ matrix acting on the tensor product space $\nu\otimes \nu$, where $N$ is the dimension of $\nu$. Such a matrix $ \check{R}$ satisfies the relativistic YBE\cite{hxg} as follows
\begin{equation}
\check{R}_{12}(u)\check{R}_{23}(\frac{u+v}{1+\beta^2uv})\check{R}_{12}(v)=
\check{R}_{23}(v)\check{R}_{12}(\frac{u+v}{1+\beta^2uv})\check{R}_{23}(u).
\end{equation}
In this paper, we focus on 3-dimensional space. The reduced relativistic YBE reads
\begin{equation}
\label{rybe}A(u)B(\frac{u+v}{1+\beta^2uv})A(v)=B(v)A(\frac{u+v}{1+\beta^2uv})B(u).
\end{equation}

Let the unitary Yang-Baxter matrix take the form
\begin{equation}
 \label{sybe}\left\{ \begin{aligned}
             &A(u)=\rho(u)(I+F(u)E_A) \\
             &B(u)=\rho(u)(I+F(u)E_B) .
         \end{aligned} \right.
\end{equation}
Following Xue \emph{et al.} \cite{wx1}, we obtain
\begin{equation}
 \left\{ \begin{aligned}
             &F(u)=\frac{e^{-2i\theta}-1}{d} ,\\
             &e^{-2i\theta}=\frac{\beta^2u^2+2i\varepsilon \beta u\sqrt{d^2/(4-d^2)}+1}
             {\beta^2u^2-2i\varepsilon \beta u\sqrt{d^2/(4-d^2)}+1} ,
         \end{aligned} \right.
\end{equation}
where the new parameter \(\theta\) is real. Let
\(\rho(u)=e^{i\theta}\). The Yang-Baxter matrix can be rewritten in
the following form
\begin{equation}
\label{sybe}\left\{ \begin{aligned}
             &A(\theta,\varphi_1,\varphi_2)=e^{i\theta}I-f(\theta)E_A ,\\
             &B(\theta,\varphi_1,\varphi_2)=e^{i\theta}I-f(\theta)E_B ,
         \end{aligned} \right.
\end{equation}
where \(f(\theta)=2i\sin{\theta}/d\).

The Yang-Baxter matrix depends on three parameters: the first is
\(\theta\) (\(\theta\) is time-independent); the others are
\(\varphi_i,(i=1,2)\) contained in the matrix \(E\). In physics the
parameter \(\varphi_1\) and \(\varphi_2\) are flux which depends on
time \(t\). Usually take \(\varphi_i=\omega_it,(i=1,2)\) and
\(\omega_i\) are the frequency. Operators $A(\theta,\varphi_1,\varphi_2)$ and $B(\theta,\varphi_1,\varphi_2)$, satisfying $B(\theta,\varphi_1,\varphi_2)=UA(\theta,\varphi_1,\varphi_2)U^{-1}$,
are unitary operators \((A(\theta,\varphi_1,\varphi_2)^{\dag}=A(\theta,\varphi_1,\varphi_2)^{-1}, B(\theta,\varphi_1,\varphi_2)^{\dag}=B(\theta,\varphi_1,\varphi_2)^{-1})\).

To simplify the following discussion, we will restrict attention to the case \(\varphi_1=-\varphi_2=\varphi\). Following Ge \emph{et al.} \cite{cxg1}, we can obtain Yang Baxter Hamiltonian through the Schr\(\ddot{o}\)dinger evolution of the states
\begin{equation}
\label{h}\hat{H}=i\hbar\frac{\partial B(\theta,\varphi)} {\partial
t}B^{\dag}(\theta,\varphi) ,
\end{equation}
where \(\varphi\) be time dependent as
\(\varphi=\omega t\) and \(\theta\) be time independent.

For convenience, we introduce the Gell-Mann matrices \(I_{\lambda}\)\cite{gm}, a
basis for \(su(3)\) algebra. Such matrices satisfy \([I_{\lambda},I_{\mu}]=i f_{\lambda \mu
\nu}I_{\nu},(\lambda,\mu,\nu=1,2,...,8)\), where $f_{\lambda \mu \nu}$ are the structure constants of $su(3)$. We denote
\(I_{\pm}=I_1\pm iI_2\), \(V_{\pm}=I_4\mp iI_5\), \(U_{\pm}=I_6\pm
iI_7\) and \(Y=\frac{2}{\sqrt{3}}I_8\). Let
\begin{equation}
\label{su2}\left\{\begin{aligned}
        S_+=&\zeta(-i(d^2-d-1)^{1/2}(e^{-i\theta}+2i\sin{\theta}d^{-2})I_++id^{1/2}(e^{-i\theta}+2i\sin{\theta}d^{-2})U_-) ,\\
        S_-=&\zeta(i(d^2-d-1)^{1/2}(e^{i\theta}-2i\sin{\theta}d^{-2})I_--id^{1/2}(e^{i\theta}-2i\sin{\theta}d^{-2})U_+) ,\\
        S_3=&\frac{1}{2}\big[(1+d-d^2)(1-d^2)^{-1}(\frac{I}{3}+\frac{Y}{2}+I_3)-(\frac{I}{3}+\frac{Y}{2}-I_3)-d(1-d^2)^{-1}(\frac{I}{3}-Y)\\
        &+d^{1/2}(d^2-d-1)^{1/2}(1-d^2)^{-1}(V_-+V_+)\big] ,
 \end{aligned}\right.
\end{equation}
where $\zeta=\frac{d^2}{\sqrt{(d^2-1)(d^4-4(d^2-1)\sin^2{\theta})}}$. These operators satisfy the \(su(2)\) algebra relations $([S_+,S_-]=2S_3, [S_3,S_{\pm}]=\pm S_{\pm}, (S_{\pm})^2=0, S_{\pm}=S_1\pm iS_2)$.

In terms of the operators\eqref{su2}, the Hamiltonian Eq.\eqref{h} can be recast as following
\begin{equation}
\begin{aligned}
\hat{H}=&-4\omega \hbar
\sin{\theta}(d^2-1)^{1/2}d^{-2}(\sin{\alpha}\cos{\beta}S_1+\sin{\alpha}\sin{\beta}S_2+\cos{\alpha}S_3).
\end{aligned}
\end{equation}
Its eigenvalues are \(E_0=0,E_{\pm}=\mp\omega \hbar \cos{\alpha} \),
where $\cos{\alpha}=\frac{2\sin{\theta}\sqrt{d^2-1}}{d^2}$, $\beta=\varphi$, here \(d\ge 1\). By the way, its Casimir operator is \(\kappa=\frac{1}{2}(S_+S_-+S_-S_+)+S_3^2\). It is easy to find the eigenvalues of \(\kappa \) are \(\frac{1}{2}(\frac{1}{2}+1)=\frac{3}{4}\) and \(0(0+1)=0\), which correspond to spin-1/2 and spin-0. According to the definition of Berry phase, when \(\varphi(t)\) evolves adiabatically from 0 to \(2\pi\), the corresponding Berry phase is
\begin{equation}
\label{bp1}\gamma_{\alpha}=i\int_0^{T}\langle
\Psi_{\alpha}|\frac{\partial }{\partial t}|\Psi_{\alpha}\rangle dt .
\end{equation}

Noting that Hamiltonian returns to its original form after the time
\(T=2\pi/\omega\), we easily obtain the corresponding Berry phases of this Yang-baxter system
\begin{equation}
 \label{sa}\left\{\begin{aligned}
        &\gamma_0=0 ,\\
        &\gamma_{\pm}=\pm\pi(1-\cos{\alpha})=\pm \frac{\Omega}{2} ,
 \end{aligned}\right.
\end{equation}
where \(\Omega=2\pi(1-\cos{\alpha})\) is the solid angle enclosed by
the loop on the Bloch sphere. The system also equals to spin-\(1/2\)
system and spin-\(0\) system.
Substituting $\cos{\alpha}=\frac{2\sin{\theta}\sqrt{d^2-1}}{d^2}$ into Eq.\eqref{sa}, we obtain $\gamma_{\pm}=\pm\pi(1-\frac{2\sin{\theta}\sqrt{d^2-1}}{d^2})$.
Substituting $\theta$ with $\frac{\pi}{2}-\theta$, we rewrite Berry phase as follows
\begin{equation}\label{bptr}
\gamma_{\pm}=\pm\pi(1-\frac{2\cos{\theta}\sqrt{d^2-1}}{d^2}).
\end{equation}

\begin{figure}
  \centering
  \subfigure[]{
    \label{fig:subfig:a} 
    \includegraphics[width=3.2in]{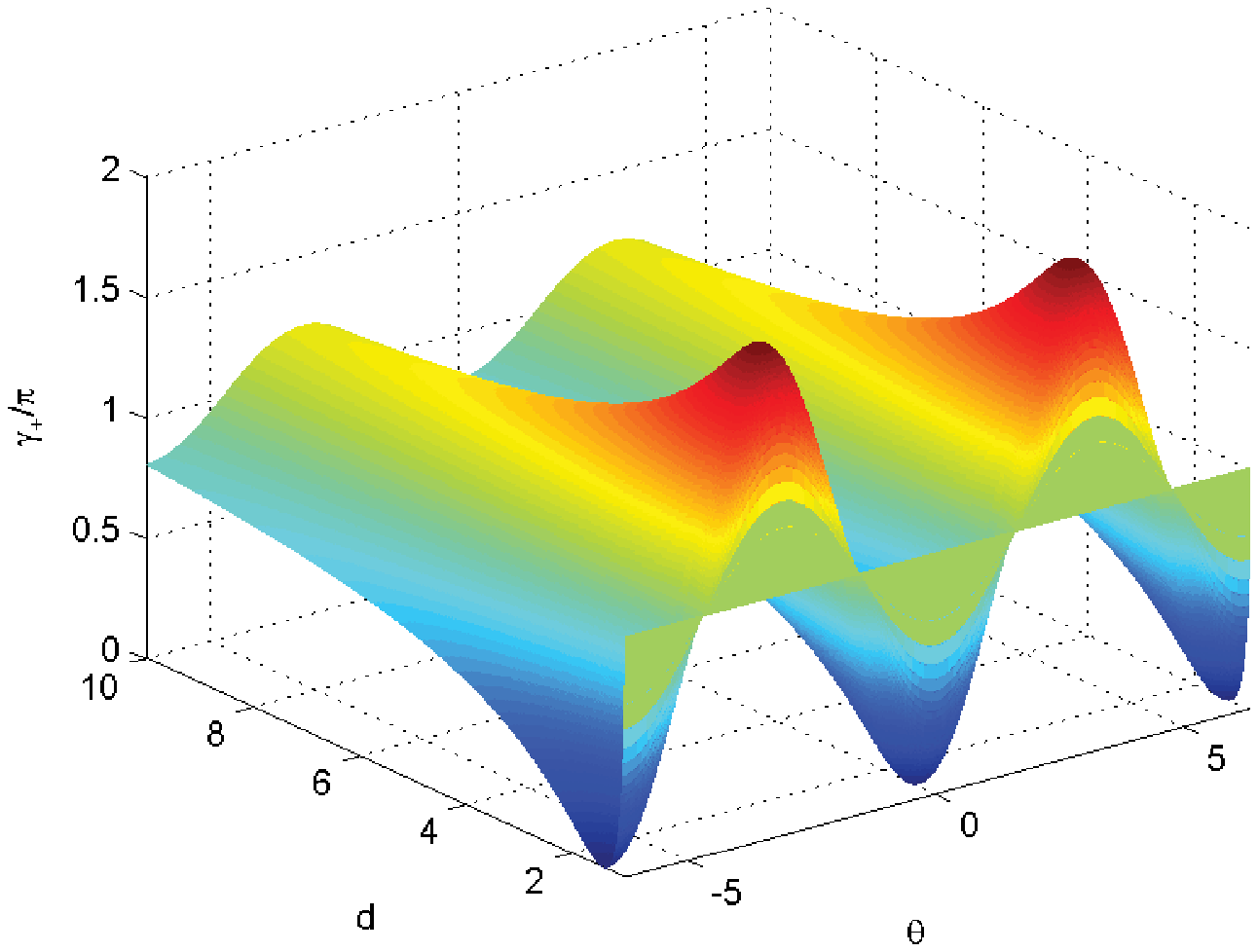}}
  \hspace{0.3in}
  \subfigure[]{
    \label{fig:subfig:b} 
    \includegraphics[width=3.2in]{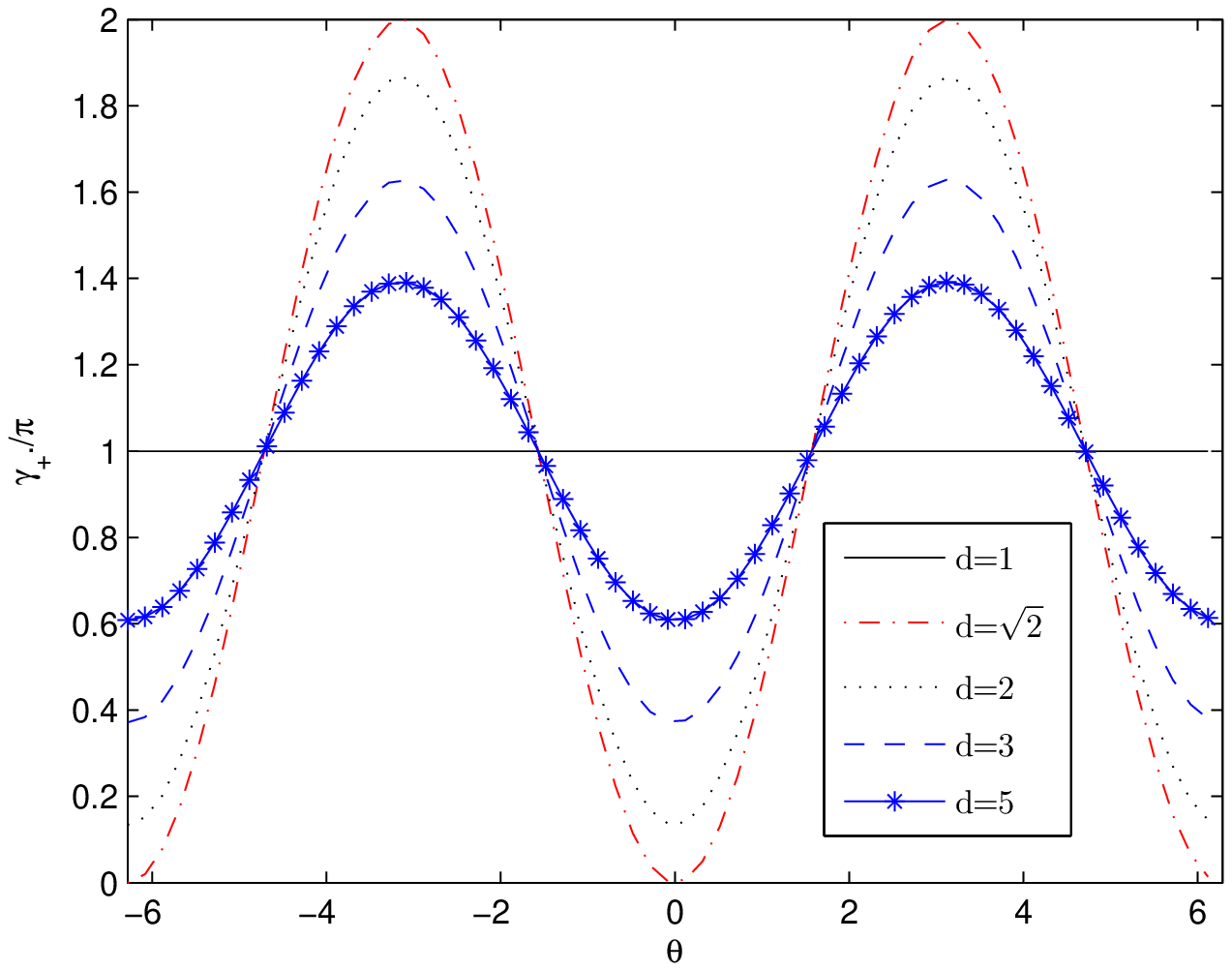}}
  \caption{(The left figure shows the Berry phase $\gamma_{+}=\pi(1-\frac{2\cos{\theta}\sqrt{d^2-1}}{d^2})$ versus the the topological parameter $d$ and the spectral parameter $\theta$. The right figure, the sectional drawings have also provided with the same values of parameters. $d=1$ (solid line), $d=\sqrt{2}$ (dot-dashed line), $d=2$ (dotted line), $d=3$ (dashed line), $d=5$ (star line).)}
  \label{fig:subfig}
\end{figure}

It is worth mentioning that in some papers\cite{cxg1}, the Berry phase $\gamma_{\pm}=\pm\pi(1-\cos{\theta})$ of Yang-Baxter system only depends on the spectral parameter $\theta$. It is interesting that in our paper, the Berry phases\eqref{bptr} not only depends on the spectral parameter $\theta$, but also depends on the topological parameter $d$. The Berry phase (Eq.\eqref{bptr}) reduce to $\gamma_{\pm}=\pm\pi(1-\cos{\theta})$ if $d=\sqrt{2}$. The FIG. \ref{fig:subfig}, which corresponds to the Berry phase $\gamma_+$. The FIG. \ref{fig:subfig:a} illustrate the Berry phases Eq.\eqref{bptr} versus the spectral parameter $\theta$ and the topological parameter $d$. The FIG. \ref{fig:subfig:b} illustrate the Berry phase $\gamma_{+}$ versus the spectral parameter $\theta$, when $d$ choice specific values. It is demonstrated that the Berry phase $\gamma_+$ is maximum (minimum) at parameter $\theta=(2n+1)\pi$ ($\theta=2n\pi$), for a given definit value of $d$. Where $n$ is integer. The maximum (minimum) of $\gamma_+$ is $\pi(1+\frac{2\sqrt{d^2-1}}{d^2})$  ($\pi(1-\frac{2\sqrt{d^2-1}}{d^2})$). As $d$ increase, the maximum of $\gamma_+$ decreases, the minimum of $\gamma_+$ increases. The Berry phase $\gamma_+$ tend to a constant value $\pi$, with $d$ approaches infinity.

\section{Summary}
In this paper we presented BWM-algebra $(A,\ B,\ E_A,\ E_B)$ and solution of YBE $(A(\theta,\varphi_1,\varphi_2),\ B(\theta,\varphi_1,\varphi_2))$ in 3-dimensional representation which satisify $B=UAU^{-1}$, $E_B=UE_AU^{-1}$ and $B(\theta,\varphi_1,\varphi_2)=UA(\theta,\varphi_1,\varphi_2)U^{-1}$. The evolution of the Yang-Baxter system is explored by constructing a Hamiltonian from the unitary $B(\theta,\varphi)$ matrix. We study the Berry phase of the Yang-Baxter system, and obtain the relationship between topological parameter and Berry phase $\gamma_{\pm}=\pm\pi(1-\frac{2\cos{\theta}\sqrt{d^2-1}}{d^2})$. Then we compare the Berry phase of Yang-Baxter system in Ref.\cite{cxg1} with us, and find the topological parameter d plays a deformation role in the Berry phase. We have been discussing in this paper is still an open problem that will require a deal of further investigations.

\section{Acknowledgments}
This work was supported in part by NSF of China (Grant No.10875026)


\begin{thebibliography}{99}

\bibitem{ye}Yang, C.N.: Some Exact Results for the Many-Body Problem in one Dimension with Repulsive Delta-Function Interaction. {\it Phys. Rev. Lett.} {\bf 19}, 1312 (1967); Yang, C.N.: S matrix for the one-dimensional N body problem with repulsive or attractive delta function interaction. {\it Phys. Rev.} {\bf 168}, 1920 (1968).
\bibitem{be}Baxter, R.J.: it Exactly Solved Models in Statistical Mechanics. New York: Academic (1982); Baxter, R.J.: Partition function of the Eight-Vertex lattice model. {\it Ann. Phys.} {\bf 70}, 193 (1972).
\bibitem{tla}Temperley, H.N.V., Lieb, E.H.: Relations between the 'Percolation' and 'Colouring' Problem and other Graph-Theoretical Problems Associated with Regular Planar Lattices: Some Exact Results for the 'Percolation' Problem. {\it Proc. Roy. Soc. London, A} {\bf  322}, 251 (1971).
\bibitem{kve}Korepin, V.E., Bogoliubov, N.M., Izergin, A.G.: Quantum Inverse Scattering Method and Correlation Functions. Cambridge University Press (1993)
\bibitem{kk}Kauffman, L.H.: Knots and Physics. Singapore: World Scientific Publ Co Ltd. (1991)
\bibitem{yg}Yang. C.N., Ge. M.L.,{\it et al.}: Braid Group, Knot Theory and Statistical Mechanics (I and II). SingaporeWorld Scientific Publ Co Ltd. (1989) and (1994).
\bibitem{bob}Baxter, R.J.: The inversion relation method for some two-dimensional exactly solved models in lattice statistics. {\it J.Stat.Phys.} {\bf 28}, 1 (1982); Owczarek, A.L., Baxter, R.J.: A Class of Interaction-Round-a-Face Models and Its Equivalence with an Ice-Type Model. {\it J.Stat.Phys.} {\bf 49}, 1093 (1987); Batchelor, M.T., Barber, M.N.: Spin-s quantum chains and Temperley-Lieb algebras. {\it J. Phys. A} {\bf 23}, L15 (1990).
\bibitem{bk}Batchelor, M.T., Kuniba, A.: Temperley-Lieb lattice models arising from quantum groups. {\it J. Phys. A} {\bf 24}, 2599 (1991).
\bibitem{yql}Li, Y.Q.: Yang Baxterization.{\it J. Math. Phys.} {\bf 34}, 2 (1993).
\bibitem{kl}Kauffman, L.H., Lomonaco Jr, S.J.: Braiding Operators are Universal Quantum Gates. {\it New J. Phys.} {\bf 6}, 413 (2004).
\bibitem{ayk}Kitaev, A.Y.:Fault-tolerant quantum computation by anyons. {\it Ann. Phys.} {\bf 303}, 2 (2003).
\bibitem{frw}Franko, J., Rowell, E.C., Wang, Z.: Extraspecial 2-Groups and Images of Braid Group Representations. {\it J. Knot Theor Ramif.} {\bf 15}, 413 (2006).
\bibitem{cxg1}Chen, J.L., Xue, K., Ge, M.L.: Braiding transformation, entanglement swapping, and Berry phase in entanglement space. {\it Phys. Rev. A.} {\bf 76}, 042324 (2007); Chen, J.L., Xue, K., Ge, M.L.: Berry phase and quantum criticality in Yang-Baxter systems. {\it Annals of Physics} {\bf 323}, 2614 (2008).
\bibitem{sw}Sun, C.F., {\it et al.}: Thermal entanglement in the two-qubit systems constructed from the Yang-Baxter R-matrix. {\it International Journal of Quantum Information} {\bf 7}, 879 (2009).
\bibitem{knot}Wadati, M., Deguchi, T., Akutsu, Y.: Exactly solvable models and knot theory. {\it Phys. Rep.} {\bf 180}, 247 (1989).
\bibitem{bwma}Birman .J, Wenzl .H.: Braids, link polynomials and a new algebra. {\it Trans. A.M.S.} {\bf 313}, 249 (1989); Murakami .J.: The Kauffman Polynomial of Links and Representation Theory. {\it Osaka J. Math.} {\bf 24}, 745 (1987).
\bibitem{tlaknot}Abramsky, S.: Temperley-Lieb Algebra: From Knot Theory to Logic and Computation via Quantum Mechanics. e-print quant-ph/0910.2737.
\bibitem{b}Berry, M.V.: Quantal phase factors accompanying adiabatic changes. {\it Proc. R. Soc. Lond. Ser. A} {\bf 392}, 45 (1984).
\bibitem{aa}Aharonov, Y., Anandan, J.: Phase change during a cyclic quantum evolution. {\it Phys. Rev. Lett.} {\bf 58}, 1593 (1987).
\bibitem{spe}Sj\"oqvist, E., Pati, A.K., Ekert, A., Anandan, J.S., Ericsson, M., Oi, D.K.L., Vedral, V.: Geometric Phases for Mixed States in Interferometry. {\it Phys. Rev. Lett.} {\bf 85}, 2845 (2000).
\bibitem{sb}Samuel, J., Bhandari, R.: General Setting for Berry's Phase. {\it Phys. Rev. Lett.} {\bf 60}, 2339 (1988).
\bibitem{ts}Tong, D.M., Sj\"oqvist, E., Kwek, L.C., Oh, C.H.: Kinematic Approach to the Mixed State Geometric Phase in Nonunitary Evolution. {\it Phys. Rev. Lett.} {\bf 93}, 080405 (2004).
\bibitem{w}Wilczek, F., Zee, A.: Appearance of Gauge Structure in Simple Dynamical Systems. {\it Phys. Rev. Lett.} {\bf 52}, 2111 (1984).
\bibitem{bpxxz}Korepin, V.E., Wu, A.C.T.: Adiabatic Transport Properties and BERRY¡¯S Phase in Heisenberg-Ising Ring. {\it International Journal of Modern Physics B} {\bf 5}, 497 (1991).
\bibitem{bmk}Appelt, S., W\(\ddot{a}\)ckerle, G., Mehring, M.: Deviation from Berry¡¯s adiabatic geometric phase in a 131Xe nuclear gyroscope. {\it Phys. Rev. Lett.} {\bf 72}, 3921 (1994).
\bibitem{jve}Jones, J., Vedral, V., Ekert, A., Castagnoli, G.: Geometric quantum computation using nuclear magnetic resonance. {\it Nature } {\bf 403}, 869 (2000).
\bibitem{dcz}Duan, L.M., Cirac, J.I., Zoller, P.: Geometric Manipulation of Trapped Ions for Quantum Computation. {\it Science} {\bf 292}, 1695 (2001).
\bibitem{ww}Wootters, W.K.: Entanglement of Formation of an Arbitrary State of Two Qubits. {\it Phys. Rev. Lett.} {\bf 80}, 2245 (1998).
\bibitem{eeh}Ekert, A., Ericsson, M., Hayden, P., Inamori, H., Jones, J.A., Oi, D.K.L., Vedral, V.: Geometric quantum computation. {\it J. Mod. Opt.} {\bf 47}, 2501 (2000).
\bibitem{gpit}Leibfried, D., {\it et al.}: Experimental demonstration of a robust, high-fidelity geometric two ion-qubit phase gate. {\it Nature} {\bf 422}, 412 (2003).
\bibitem{bpss}Leek, P.J., {\it et al.}: Observation of Berry's Phase in a Solid-State Qubit. {\it science} {\bf 318}, 1889 (2007).
\bibitem{hxg}Hu, S.W., Xue, K., Ge, M.L.£ºOptical simulation of the Yang-Baxter equation. {\it Phys. Rev. A} {\bf 78}, 022319 (2008).
\bibitem{tqc1}Nayak, C., Simon, S.H., Stern, A., Freedman, M., Sarma, S.D.£ºNon-Abelian anyons and topological quantum computation. {\it Rev. Mod. Phys.} {\bf 80}, 1083 (2008).
\bibitem{tqc2}Hikami, K.: Skein theory and topological quantum registers: Braiding matrices and topological entanglement entropy of non-Abelian quantum Hall states. {\it Ann. Phys.} {\bf 323}, 1729 (1987).
\bibitem{cgx}Cheng, Y.,Ge, M.L., Xue, K.: Yang-Baxterization of braid group representations. {\it Commun Math Phys.} {\bf 136}, 195 (1991).
\bibitem{gx2}Ge, M.L., Xue, K.: Trigonometric Yang-Baxterization of colored R-matrix. {\it J. Phys. A: Math.} {\bf 26}, 281 (1993).
\bibitem{j}Jones, V.F.R.: On a certain value of the Kauffman polynomial. {\it Commun Math Phys.} {\bf 125}, 459 (1987).
\bibitem{wx1}Wang, G.C., {\it et al.}: Temperley-Lieb algebra, Yang-Baxterization and universal gate. {\it Quantum Information Processing.} {\bf 9}, 699 (2009).
\bibitem{gm}Pfeifer, W.: The Lie Algebras \(SU(N)\), An Introduction. Birkhauser Verlag (2003).




\end{thebibliography}
\end{document}